\documentclass{ws-procs9x6}

\begin{document}

\title{Neutrino Physics: A Selective Overview
\footnote{\uppercase{P}roceedings of the \uppercase{L}ake
\uppercase{L}ouise \uppercase{W}inter \uppercase{I}nstitute 2006.
\uppercase{S}lides available at http://www.phas.ubc.ca/$\sim$oser/
\uppercase{D}ue to length restrictions \uppercase{I} have been forced
to be selective and emphasize only recent results, and apologize to
the many excellent researchers whose work has been neglected as a
result.  }}

\author{SCOTT~M. OSER}

\address{University of British Columbia \\
Department of Physics \& Astronomy\\
6224 Agricultural Road \\
Vancouver BC  V6T~1Z1, Canada\\
E-mail: oser@phas.ubc.ca}  

\maketitle

\abstracts{Neutrinos in the Standard Model of particle physics are
  massless, neutral fermions that seemingly do little more than
  conserve 4-momentum, angular momentum, lepton number, and lepton
  flavour in weak interactions.  In the last decade conclusive
  evidence has demonstrated that the Standard Model's
  description of neutrinos does not match reality.  We now know that
  neutrinos undergo flavour oscillations, violating lepton flavour
  conservation and implying that neutrinos have non-zero mass.  A rich
  oscillation phenomenology then becomes possible, including
  matter-enhanced oscillation and possibly CP violation in the
  neutrino sector.  Extending the Standard Model to include neutrino
  masses requires the addition of new fields and mass terms, and
  possibly new methods of mass generation.  In this review article I
  will discuss the evidence that has established the existence of
  neutrino oscillation, and then highlight unresolved issues in
  neutrino physics, such as the nature of three-generational mixing
  (including CP-violating effects), the origins of neutrino mass, the
  possible existence of light sterile neutrinos, and the difficult
  question of measuring the absolute mass scale of neutrinos.}

\section{Neutrinos In The Standard Model}

A neutrino can be defined as a chargeless, colourless fermion.  As
such, neutrinos have only weak interactions, with tiny cross-sections,
and are exceedingly difficult to detect.  In the Standard Model of
particle physics, there is one massless neutrino associated with each
charged lepton ($e$, $\mu$, or $\tau$), and lepton flavour is
rigorously conserved, so that for example the total number of
``electron''-type leptons (charged or otherwise) is unchanged in all
interactions.  Indeed, an electron neutrino can be defined simply as
the kind of neutrino produced when a $W$ particle couples to an
electron.  Weak interactions are never observed to couple a charged
lepton $\ell$ to the wrong type of neutrino.  Nor do neutral
current ($Z$-mediated) interactions couple together two
neutrinos of different flavours.  Interestingly, although no Standard
Model process violates lepton flavour number, there is no associated
symmetry of the Lagrangian that requires this to be so---that is, the
absence of lepton-flavour-changing terms in the Lagrangian seems to be
``accidental'', and not the result of a deeper symmetry.

One of the most characteristic features of neutrinos in the Standard
Model is that weak interactions couple only to left-handed neutrinos,
or to right-handed antineutrinos.  That is, in all cases the spin of a
(massless) neutrino is observed to be antiparallel to its direction of motion.
This characteristic is associated with the $V$-$A$ nature of weak
interactions.  Whereas the electromagnetic current of an electron is
given by 
\begin{eqnarray}
j^\mu_{EM} \propto \bar{e}\gamma^\mu e 
\end{eqnarray}
the weak current that couples a $\nu_e$ to an electron has the form
\begin{eqnarray}
j^\mu_{weak} \propto \bar{e}\gamma^\mu(1-\gamma^5) \nu_e.
\end{eqnarray}
The presence of the $1-\gamma^5$ factor (a $V$-$A$ term) in the
current projects out the left-handed chirality component of the
$\nu_e$.  The result is that weak interactions only couple to
left-handed neutrino states.

The failure to observe right-handed neutrinos suggests a plausibility
argument as to why neutrinos could be expected to be massless.  The
apparent absence of right-handed neutrinos implies either that no
$\nu_R \equiv (1+\gamma^5)\nu$ state exists, or if a $\nu_R$ does
exist, then it happens to be a ``sterile'' state, having no couplings
to any vector gauge bosons.  Rather than postulate the existence of a
$\nu_R$ state that has never been seen and lacks even weak
interactions, an appeal to Ockham's razor suggests the more economical
solution that the right-handed field $\nu_R$ not exist at all.
However, in the Standard Model, a mass term is a term in the
Lagrangian that couples left-handed and right-handed states:
\begin{eqnarray}
\mathcal{L} = - m \bar{\psi}\psi = -m(\bar{\psi_L}\psi_R +
\bar{\psi_R}{\psi_L})   
\end{eqnarray}
Accordingly, if no $\nu_R$ exists, then one cannot form such a mass
term, and so the neutrino must be massless.  The alternative is
seemingly to postulate the existence of right-handed neutrino states
which don't participate in even weak interactions but which provide
the fields needed to produce neutrino masses.  This unpalatable
situation, as much as the fact that experimentally neutrino masses
turned out to be immeasurably small, provided justification for
assuming the neutrino mass to be zero in the Standard Model.  That
assumption turns out to be wrong, but is less of an {\em ad hoc}
assumption than is sometimes claimed when one keeps in mind that the
simplest alternative forces us to introduce sterile fermion fields
even more ethereal that the neutrino itself!

\section{Phenomenology Of Neutrino Oscillation}

The Standard Model neutrinos strike me as rather dismal particles in
the end.  With no mass and very limited interactions, the major
practical import that neutrinos seem to have is to provide a ``junk''
particle to balance a number of conservation laws such as 4-momentum,
angular momentum, lepton number, and lepton flavour.  Given this
situation, and the difficulties associated with neutrino experiments
to begin with, it is perhaps not surprising that neutrinos were for
a long time a neglected area of particle phenomenology.  

Some progress was made in 1962 when Maki, Nakagawa, and Sakata
proposed (in true theorist fashion, on the basis of zero experimental
evidence) a new phenomenon now known as neutrino
oscillation.\cite{mns}  The inspiration for this proposal was the
observation that charged current interactions on quarks produce
couplings between quark generations.  For example, while naively we
would expect the interactions of a $W^\pm$ to couple $u$ to $d$, $s$
to $c$, or $t$ to $b$, weak decays such as $\Lambda^0 \to p\pi^-$ are
also observed in which an $s$ quark gets turned into a $u$ quark, thus
mixing between the second and first quark generations.  We describe
this by saying that there is a rotation between the mass eigenstates
(e.g. $u$,$d$,$s$ ...) produced in strong interactions, and the weak
eigenstates that couple to a $W$ boson.  In this language, a $W$ does
not simply couple a $u$ quark to a $d$ quark, but rather it couples
$u$ to something we can call $d'$, which is a linear superposition of
the $d$, $s$, and $b$ quarks.  We describe this ``rotation'' between
the strong and weak eigenstates by a 3$\times$3 unitary matrix called
the CKM matrix:
\begin{eqnarray}
\left( \begin{array}{c}
d' \\
s' \\
b' \\
\end{array} \right) =
\left[ \begin{array}{ccc}
V_{ud} & V_{us} & V_{ub} \\
V_{cd} & V_{cs} & V_{cb} \\
V_{td} & V_{ts} & V_{tb} \\
\end{array} \right]
\left( \begin{array}{c}
d \\
s \\
b \\
\end{array} \right)
\end{eqnarray}
The off-diagonal elements of this matrix allow transitions between
quark generations in charged current weak interactions, and through a
complex phase in matrix $V$ also produce CP violation in the quark
sector.  The measurement of the CKM matrix elements and exploration of
its phenomenology has been one of the most active fields in particle
physics for the past four decades.

Maki, Nakagawa, and Sakata (hereafter known as MNS) proposed that
something similar could happen in the neutrino sector.\cite{mns}
Once the muon neutrino was discovered in 1962\cite{munu}, it became
possible to suppose that neutrino flavour eigenstates such as $\nu_e$
or $\nu_\mu$ might not correspond to the neutrino mass eigenstates.
That is, the particle we call ``$\nu_e$'', produced when an electron
couples to a $W$, might actually be a linear superposition of two mass
eigenstates $\nu_1$ and $\nu_2$.  In the case of 2-flavour mixing, we
can write:
\begin{eqnarray}
\left( \begin{array}{c}
\nu_e \\
\nu_\mu \\
\end{array} \right) =
\left[ \begin{array}{cc}
+\cos \theta & ~~+\sin \theta \\
-\sin \theta & ~~+\cos \theta \\
\end{array} \right]
\left( \begin{array}{c}
\nu_1 \\
\nu_2 \\
\end{array} \right)
\label{eq:2x2}
\end{eqnarray}
While the formalism is exactly parallel to that used for quark mixing,
with angle $\theta$ in Equation~\ref{eq:2x2} playing the role of a
Cabibbo angle for leptons, the resulting phenomenology is somewhat
different.  In the case of quarks, mixing between generations can be
readily seen by producing hadrons through strong interactions, and
then observing their decays by weak interactions.  For example, we can
produce a $K^+$ in a strong interaction, then immediately observe the
decay $K^+ \to \pi^0 e^+ \nu_e$, in which an $\bar{s}$ turns into a
$\bar{u}$.  Neutrinos, however, have {\em only} weak interactions, and
so we cannot do the trick of producing neutrinos by one kind of
interaction and then detecting them with a different interaction.  In
other words, a rotation between neutrino flavour eigenstates and
neutrino mass eigenstates such as in Equation~\ref{eq:2x2} has no
direct impact on weak interaction vertices themselves.  $W$ bosons
will still always couple an $e$ to a $\nu_e$ and a $\mu$ to $\nu_\mu$
even if there is a rotation between the flavour and mass eigenstates.

\begin{figure}[ht]
\centerline{\epsfxsize=3.5in\epsfbox{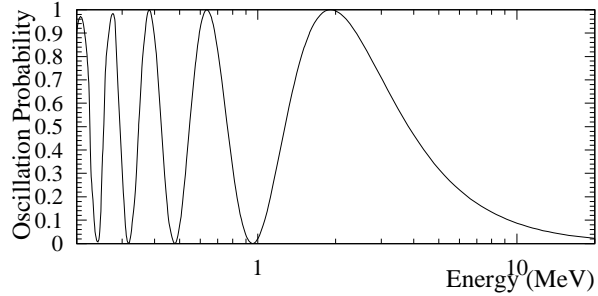}}   
\caption{Oscillation probability as a function of neutrino energy for
  a fixed value of $\Delta m^2 L$, with $\sin^2 2\theta=1$.
\label{fig:energyosc}}
\end{figure}

To observe the effects of neutrino mixing we therefore must resort to
some process that depends on the properties of the mass eigenstates.
While the flavour basis is what matters for weak interactions, the
mass eigenstate is actually what determines how neutrinos propagate as
free particles in a vacuum.  Imagine, for example, that we produce at
time $t=0$ a $\nu_e$ state with some momentum $\vec{p}$:
\begin{eqnarray}
 |\nu_e (t=0) \rangle = \cos \theta |\nu_1 \rangle + \sin \theta |
  \nu_2 \rangle 
\end{eqnarray}
As this state propagates in vacuum, each term picks up the standard
quantum mechanical phase factor for plane wave propagation:
\begin{eqnarray}
 |\nu (\vec{x},t) \rangle = \exp(i(\vec{p}\cdot \vec{x} - E_1t))
\cos \theta |\nu_1 \rangle + \exp(i(\vec{p}\cdot \vec{x} - E_2t))
\sin \theta |  \nu_2 \rangle 
\end{eqnarray}
Here the energy $E_i$ of the $i$th mass eigenstate is given by the
relativistic formula $E_i = \sqrt{\vec{p}^2 + m_i^2}$, and $\hbar
\equiv c \equiv 1$.  If the two mass eigenstates $\nu_1$ and $\nu_2$
have identical masses, then the two components will have identical
momenta and energy, and so share a common phase factor of no physical
significance.  However, suppose that $m_1 \ne m_2$.  If $m_i \ll p
\equiv |\vec{p}|$, then we can expand the formula for $E_i$ as
follows:
  \begin{eqnarray}
 E_i = \sqrt{p^2 + m_i^2} = p\sqrt{1+m_i^2/p^2} \approx p + m^2_i/(2p)
  \end{eqnarray}
At some time $t>0$, the neutrino's state will be proportional to the
following superposition:
\begin{eqnarray}
|\nu(t) \rangle \propto \cos \theta |\nu_1 \rangle
+ e^{i\phi} \sin \theta |\nu_2\rangle  
\end{eqnarray}
with the phase difference $\phi$ being given by 
\begin{eqnarray}
\phi = \left( \frac{m_1^2}{2p} - \frac{m_2^2}{2p} \right) t
\label{eq:phi}
\end{eqnarray}
The net result is that at time $t$, the neutrino that originally was
in a pure $\nu_e$ state is no longer in a pure $\nu_e$ state, but due
to the phase difference $\phi$ will have acquired a non-zero component
of $\nu_\mu$!  We therefore can determine the
probability that our original $\nu_e$ will interact as a $\nu_\mu$,
which by Equation~\ref{eq:phi} depends on $\Delta m^2 \equiv m_1^2 -
m_2^2$, $p \approx E$, and $t \approx L/c$ in the relativistic limit:
\begin{eqnarray}
P(\nu_e \to \nu_\mu) =
| \langle \nu_\mu | \nu(t) \rangle |^2 =
\sin^22\theta \sin^2 \left( \frac{1.27 \Delta m^2 L}{E} \right)
\label{eq:vacuum}
\end{eqnarray}
In this formula $\Delta m^2$ is given in eV$^2$, $L$ is the distance the
neutrino has travelled in km, and $E$ is the neutrino energy in GeV.
The oscillation probability in Equation~\ref{eq:vacuum} has a
characteristic dependence on both $L$ and $E$ that is a distinctive
signature of neutrino oscillations.  Figure~\ref{fig:energyosc} shows
the oscillation probability vs. energy for representative parameters.

While Equation~\ref{eq:vacuum} suffices to describe oscillations
involving two neutrino flavours in vacuum, the presence of matter
alters the neutrino propagation, and hence the oscillation
probability.\cite{msw}  The reason for this is that ordinary matter is
flavour-asymmetric.  In particular, normal matter contains copious
quantities of electrons, but essentially never any $\mu$'s or $\tau$'s.
As a result, $\nu_e$'s travelling through matter can interact with
leptons in matter by both $W$ and $Z$ boson exchange, while $\nu_\mu$ or
$\nu_\tau$ can interact only by $Z$ exchange.  This difference affects
the amplitude for forward scattering (scattering in which no momentum
is transferred).  Electron neutrinos pick up an extra interaction
term, proportional to the density of electrons in matter, that acts as
a matter-induced potential that is different for $\nu_e$'s than for
other flavours.  Effectively $\nu_e$'s travelling through matter have
a different ``index of refraction'' than the other flavours.
Equation~\ref{eq:msw} shows the time evolution of the neutrino flavour
in the flavour basis including both mixing and the
matter-induced potential:
\begin{equation}
i\frac{d}{dt} \left( \begin{array}{c}
\nu_e  \\
\nu_\mu 		     \end{array} \right) = 
\left(
\begin{array}{ll}
-\frac{\Delta m^2}{4E} \cos 2\theta + \sqrt{2}G_FN_e~~ & 
\frac{\Delta  m^2}{4E} \sin 2\theta \\
\frac{\Delta  m^2}{4E} \sin 2\theta & \frac{\Delta  m^2}{4E} \cos 2\theta \\
\end{array} \right) \left(
\begin{array}{c}
  \nu_e \\ \nu_\mu
\end{array}
 \right) 
\label{eq:msw}
\end{equation}
The additional term $\sqrt{2}G_FN_e$ appearing in
Equation~\ref{eq:msw} is the matter-induced potential, which is
proportional to the electron number density $N_e$ and is linear in
$G_F$.  This effect, known as the MSW effect after Mikheev, Smirnov,
and Wolfenstein\cite{msw}, gives rise to a rich phenomenology in which
oscillation probabilities in dense matter, such as the interior of the
Sun, can be markedly different from those seen in vacuum.  Of the
experimental results to date, only in solar neutrino oscillations does
the MSW effect play a significant role, although future long-baseline
neutrino oscillation experiments also may have some sensitivity to
matter effects.

The generalization of neutrino mixing and oscillation to three
flavours is straightforward.  Instead of a 2$\times$2 mixing matrix,
as in Equation~\ref{eq:2x2}, we relate the neutrino flavour
eigenstates to the neutrino mass eigenstates by a 3$\times$3 unitary
matrix, completely analogous to the CKM matrix for quarks.  The
neutrino mixing matrix is known as the MNS matrix for Maki, Nakagawa,
and Sakata, and occasionally as the PMNS matrix when acknowledging
Pontecorvo's early contributions to the theory of neutrino
oscillations.\cite{mns}
\begin{equation}
\begin{array}{r}
\left( \begin{array}{c}
\nu_e \\
\nu_\mu \\
\nu_\tau \\
\end{array} \right) =
\left[ \begin{array}{ccc}
U_{e 1} & U_{e 2} & U_{e 3} \\
U_{\mu 1} & U_{\mu 2} & U_{\mu 3} \\
U_{\tau 1} & U_{\tau 3} & U_{\tau 3} \\
\end{array} \right]
\left( \begin{array}{c}
\nu_1 \\
\nu_2 \\
\nu_3 \\
\end{array} \right) \\
\\
\approx
\left[ \begin{array}{ccc}
\phantom{-}0.9\phantom{5} & \phantom{-}0.5 & ~U_{e3} \\
-0.35 & \phantom{-}0.6 & 0.7 \\
\phantom{-}0.35 & -0.6 & 0.7 \\
\end{array} \right]
\left( \begin{array}{c}
\nu_1 \\
\nu_2 \\
\nu_3 \\
\end{array} \right)\\
\end{array}
\label{eq:mns}
\end{equation}
Equation~\ref{eq:mns} gives the approximate values of the MNS matrix
elements.  The values of all of the elements except $U_{e3}$ have been
inferred at least approximately.  The most striking feature of the MNS
matrix is how utterly non-diagonal it is, in marked contrast to the CKM
matrix.  Neutrino mixings are in general large, and there is not even
an approximate correspondence between any mass eigenstate and any flavour
eigenstate.  (Therefore it really does not make any sense to talk even
approximately about the ``mass'' of a $\nu_e$, except as a weighted
average of its constituent mass eigenstates.)  Only the unknown
matrix element $U_{e3}$ is observed to be small, with a current upper limit
of $|U_{e3}|^2 < 0.03$ (90\% confidence limit).\cite{chooz}
Section~\ref{sec:evidence} will enumerate the many lines of evidence
that demonstrate that neutrinos do in fact oscillate, and describe how
the mixing parameters are derived.

\section{Evidence For Neutrino Flavour Oscillation}
\label{sec:evidence}

Since 1998 conclusive evidence has been found demonstrating neutrino
flavour oscillation of both atmospheric neutrinos and solar
neutrinos.\cite{superk_atmos98,sno_general}  In each case the oscillation
effects have been confirmed by followup experiments using man-made
sources of neutrinos.\cite{kamland,k2k}  Here I review the
experimental situation, with a strong bias towards recent results.

\subsection{The Solar Neutrino Problem, With Solution}
The earliest indications of neutrino oscillations came from
experiments designed to measure the flux of neutrinos produced by the
nuclear fusion reactions that power the Sun.  The Sun is a prolific
source of $\nu_e$'s with energies in the $\sim$0.1-20~MeV range,
produced by the fusion reaction
\begin{equation}
4p + 2e^- \rightarrow ^4{\rm He} ~+~ 2\nu_e ~+~ 26.731~{\rm MeV.}
\label{eq:fusion}
\end{equation}


The reaction in Equation~\ref{eq:fusion} actually proceeds through a
chain of sub-reactions called the $pp$ chain, consisting of several
steps.\cite{bahcall}  Each neutrino-producing reaction in the $pp$
chain produces a characteristic neutrino energy spectrum that depends
only on the underlying nuclear physics, while the rates of the
reactions must be calculated through detailed astrophysical models of
the Sun.  Experimentally the $pp$, $^8$B, and $^7$Be reactions are
the most important neutrino-producing steps of the $pp$ chain.

The pioneering solar neutrino experiment was Ray Davis's chlorine
experiment in the Homestake mine near Lead, South Dakota.\cite{davis}
This experiment measured solar neutrinos by observing the rate of Ar
atom production through the reaction $\nu_e + ^{37}$Cl$\to
^{37}$Ar$+e^-$.  By placing 600 tons of tetrachloroethylene deep
underground (to shield it from surface radiation), and using
radiochemistry techniques to periodically extract and count the number
of argon atoms in the tank, Davis inferred a solar neutrino flux that
was just $\sim$1/3 of that predicted by solar model
calculations.\cite{davis,ssm}

This striking discrepancy between theory and experiment at first had
no obvious particle physics implications.  Both the inherent
difficulty of looking for a few dozen argon atoms inside 600 tons of
cleaning fluid, and skepticism about the reliability of solar model
predictions, cast doubt upon the significance of the disagreement.  A
further complication is that the reaction that Davis used to measure
the $\nu_e$ flux was sensitive to multiple neutrino-producing
reactions in the $pp$ chain, making it impossible to determine which
reactions in the Sun are not putting out enough neutrinos.

When scrutiny of both the Davis experiment and the solar model
calculations failed to uncover any clear errors, other experiments
were built to measure solar neutrinos in other ways.  The Kamiokande
and Super-Kamiokande water Cherenkov experiments have measured elastic
scattering of electrons by $^8$B solar neutrinos, using the
directionality of the scattered electrons to confirm that the
neutrinos in fact are coming from the Sun.\cite{sk_solar}  The
measured elastic scattering rate is just $\sim$47\% of
the solar model prediction.  The SAGE and
GNO/GALLEX experiments have employed a different radiochemical
technique to observe the $\nu_e + ^{71}$Ge$\to ^{71}$Ge$+e^-$
reaction, which is primarily sensitive to $pp$ neutrinos, and have
measured a rate that is $\sim$55\% of the solar model
prediction.\cite{gallium}

Multiple experiments using different techniques have therefore
confirmed a deficit of solar $\nu_e$'s relative to the model
predictions.  Although interpretation of the data is complicated by
the fact that each kind of experiment is sensitive to neutrinos of
different energies produced by different reactions in the $pp$ fusion
chain, in fact there is apparently no self-consistent way to modify
the solar model predictions that will bring the astrophysical
predictions into agreement with the experimental results.  This
situation suggested that the explanation of the solar neutrino problem
may not lie in novel astrophysics, but rather might indicate a problem
with our understanding of neutrinos.

While it was realized early on that neutrino oscillations that
converted solar $\nu_e$ to other flavours (to which the various
experiments wouldn't be sensitive) could explain the observed
deficits, merely observing deficits in the overall rate was generally
considered insufficient grounds upon which to establish neutrino
oscillation as a real phenomenon.  It was left for the Sudbury
Neutrino Observatory (SNO) to provide the conclusive evidence that
solar neutrinos change flavour by directly counting the rate of all
active neutrino flavours, not just the $\nu_e$ rate to which the other
experiments were primarily sensitive. 

SNO is a water Cherenkov detector that uses 1000 tonnes of D$_2$O as
the target material.\cite{sno_nim}  Solar neutrinos can interact with
the heavy water by three different interactions:
\begin{equation}
\begin{array}{llll}
(CC)~~~~~~~ & \nu_e + d & \to & p + p + e^- \\  
(NC) & \nu_x + d & \to & p + n + \nu_x \\  
(ES) & \nu_x + e^- & \to & \nu_x + e^- \\  
\end{array}
\end{equation}
Here $\nu_x$ is any active neutrino species. The reaction thresholds
are such that SNO is only sensitive to $^8$B solar
neutrinos.\footnote{The tiny flux of higher-energy neutrinos from the
  $hep$ chain may be neglected here.}  The
charged current (CC) interaction measures the flux of $\nu_e$'s coming
from the Sun, while the neutral current (NC) reaction measures the
flux of all active flavours.  The elastic scattering (ES) reaction is
primarily sensitive to $\nu_e$, but $\nu_\mu$ or $\nu_\tau$ also
elastically scatter electrons with $\sim 1/6$th the cross section of
$\nu_e$.  

SNO has measured the effective flux of $^8$B neutrinos inferred from
each reaction.  In units of $10^6$~neutrinos/cm$^2$/s the most recent
measurements are\cite{sno_nsp}:
\begin{equation}
\begin{array}{rl}
\phi_{CC} = & 1.68 \pm 0.06~{\rm (stat.)} ^{+0.08}_{-0.09}~ {\rm (sys.)}
\\
\phi_{NC} = &  4.94 \pm0.21~{\rm (stat.)} ^{+0.38}_{-0.34}~ {\rm (sys.)} \\
\phi_{ES} = & 2.34 \pm 0.22~{\rm (stat.)} ^{+0.15}_{-0.15}~ {\rm (sys.)}
\\
\end{array}
\end{equation}
In short, the NC flux is found to be in good agreement with the solar
model predictions, while the CC and ES rates are each consistent with
just $\sim35\%$ of the $^8$B flux being in the form of $\nu_e$'s.  

This direct demonstration that $\phi_e < \phi_{total}$ provides
dramatic proof that solar neutrinos change flavour, resolving the
decades-old solar neutrino problem in favour of new neutrino physics.
The neutrino oscillation model gives an excellent fit to the
data from the various solar experiments, with mixing parameters of
$\Delta m^2 \approx 10^{-4}-10^{-5}$~eV$^2$ and $\tan^2 \theta \approx
0.4-0.5$.  This region of parameter space is called the Large Mixing
Angle solution to the solar neutrino problem.  In this region of
parameter space, the MSW effect plays a dominant role in the
oscillation, and in fact $^8$B neutrinos are emitted from the Sun in
an almost pure $\nu_2$ mass eigenstate.

\subsection{KamLAND}

Although neutrino oscillations with an MSW effect are the most
straightforward explanation for the observed flavour change of solar
neutrinos, the solar data by itself cannot exclude more exotic
mechanisms of inducing flavour transformation.  However, additional
confirmation of solar neutrino oscillation has recently come from an
unlikely terrestrial experiment called KamLAND.

\begin{figure}[t]
\centerline{\epsfxsize=3.6in\epsfbox{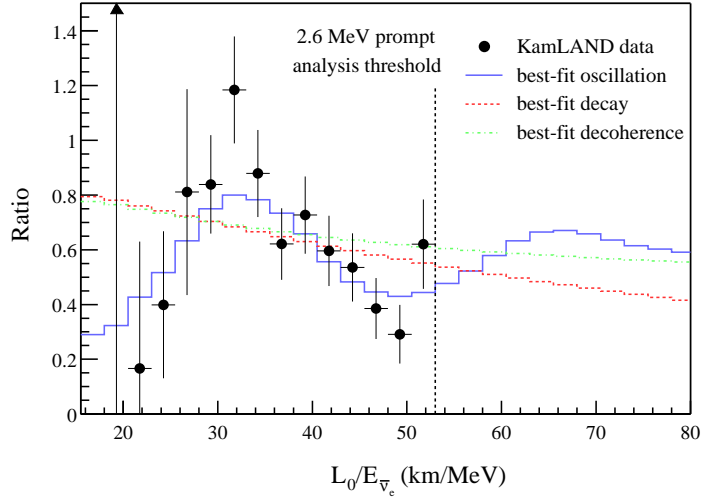}}   
\caption{Ratio of the $\bar{\nu}_e$ reactor antineutrino flux measured
  by KamLAND to the expected flux without oscillations, as a function of $L/E$.
\label{fig:kamland}}
\end{figure}

KamLAND is an experiment in Japan that counts the rate of
$\bar{\nu}_e$ produced in nuclear reactors throughout central
Japan.\cite{kamland}  If neutrinos really do oscillate with parameters
in the LMA region, then the standard oscillation theory predicts that
reactor $\bar{\nu}_e$'s, with a peak energy of $\sim3$~MeV, should
undergo vacuum oscillations over a distance of $\sim
200$~km.\footnote{At these low energies matter effects inside the
Earth are negligible.}  By integrating the flux from multiple
reactors, KamLAND achieves sensitivity to this effect.
Figure~\ref{fig:kamland} shows the $L/E$ dependence of the measured
reactor $\bar{\nu}_e$ flux divided by the expected flux at
KamLAND.\cite{kamland}  The observed flux is lower than the ``no
oscillation'' expectation on average by $\sim$1/3, with an
energy-dependent suppression of the $\bar{\nu}_e$ flux.  The pattern
of the flux suppression is in good agreement with the neutrino
oscillation hypothesis with oscillation parameters in the LMA region.

That KamLAND observes an energy-dependent suppression of the reactor
$\bar{\nu}_e$ flux, just as predicted by fits of the oscillation model
to solar neutrino data, is dramatic confirmation of the solar neutrino
results and demonstrates that neutrino oscillation is the correct
explanation of the flavour change of solar neutrinos observed by the
SNO experiment.

\begin{figure}[ht]
\centerline{\epsfxsize=3.1in\epsfbox{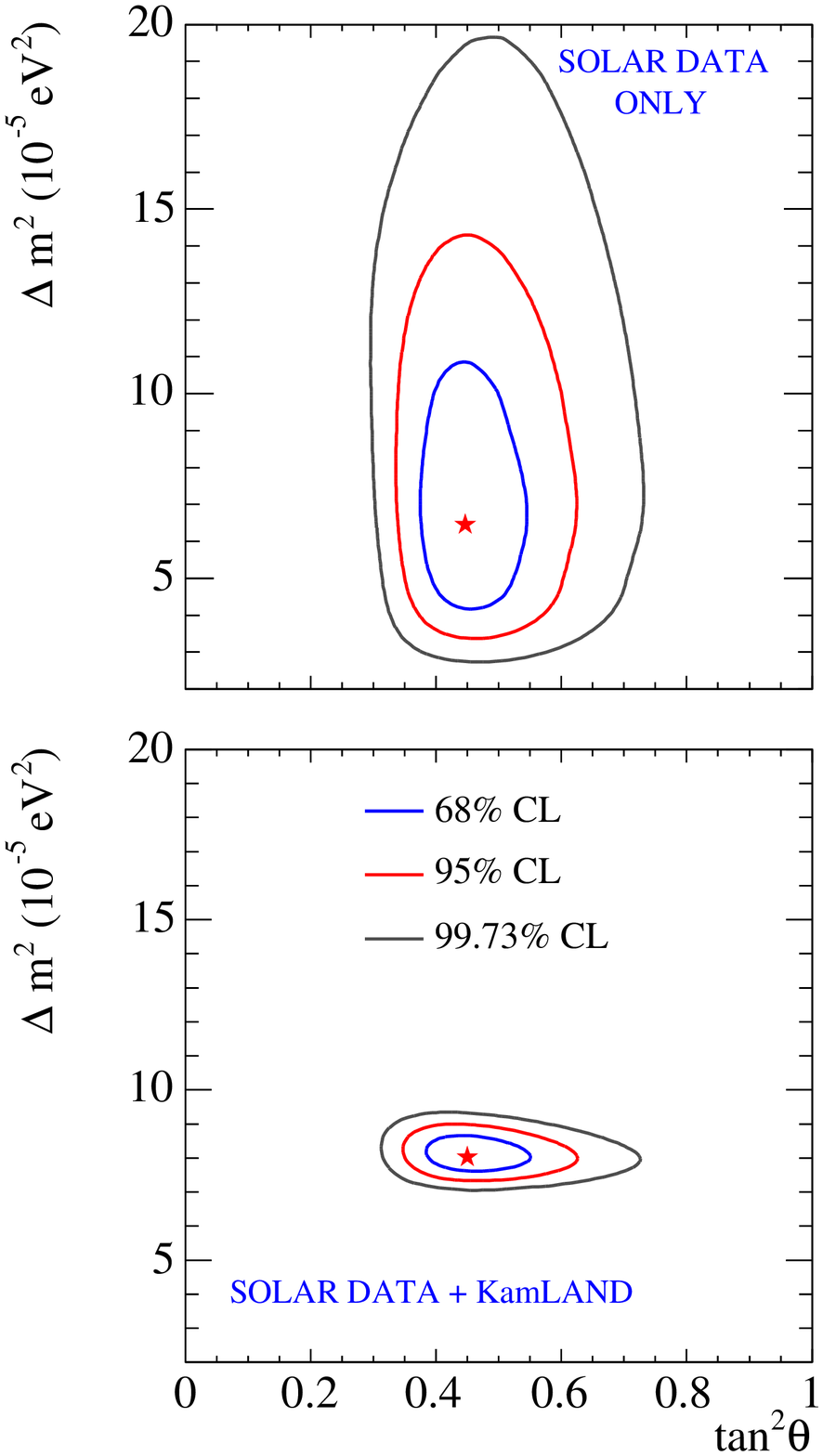}}   
\caption{Oscillation parameter contours for solar neutrino data (top),
  and solar data + KamLAND data (bottom).
\label{fig:solarmixing}}
\end{figure}

The solar experiments and KamLAND provide complementary constraints on
the mixing parameters.  Figure~\ref{fig:solarmixing} demonstrates that
solar neutrino experiments provide reasonably tight constraints on the
mixing parameter $\tan^2 \theta$, while the addition of KamLAND data
sharply constrains the $\Delta m^2$ value.\cite{sno_nsp}  This is because in
the LMA region the solar neutrino survival probability determines the
mixing angle through
\begin{equation}
|U_{e2}|^2 \approx \sin^2 \theta_{12} \approx \frac{\phi_{CC}}{\phi_{NC}}  
\end{equation}
while the observation of a distortion in the reactor antineutrino
energy spectrum fixes $\Delta m_{21}^2$.  Here the subscripts on
$\theta_{12}$ and $\Delta m_{21}^2$ reflect the fact that solar
neutrino oscillations involve the first and second mass eigenstates.

\subsection{Atmospheric Neutrinos}

Although the solar neutrino problem provided early indications
that the Standard Model's description of neutrinos is incomplete,
resolution of the solar neutrino problem was a long time coming, and
the first {\em conclusive} demonstration of neutrino oscillation
actually came from studies of atmospheric neutrinos.  Atmospheric
neutrinos are produced when cosmic rays (primarily protons) collide in
the upper atmosphere to make hadronic showers.  These showers contain
charged pions, which decay leptonically by $\pi^\pm \to \mu^\pm
\nu_\mu$.  The muons in turn generally decay in flight by $\mu^\pm \to
e^\pm \nu_\mu \nu_e$, where I've ignored differences between $\nu$ and
$\bar{\nu}$ states.  A robust conclusion that follows from the decay
sequence is that the ratio of $\nu_\mu$ to $\nu_e$ in the atmospheric
neutrino flux should be 2:1.

In 1998 the Super-Kamiokande collaboration reported results showing
that that ratio of the flux of $\nu_\mu$ to $\nu_e$ in fact is not
2:1, but is closer to 1:1.\cite{superk_atmos98,sk_atmos}  Closer examination
revealed that while the $\nu_e$ flux in fact is in good agreement with
Monte Carlo predictions, the $\nu_\mu$ flux shows a marked deficit.
The size of this deficit varies with neutrino energy, and with the zenith
angle of the event.  This latter point is significant in that
downgoing neutrinos are produced in the atmosphere just overhead, and
have travelled $<10$~km before reaching Super-Kamiokande, while
upgoing neutrinos are produced in the atmosphere on the far side of
the Earth, and have travelled $\sim$13,000~km before reaching the
detector.
\begin{figure}[ht]
\centerline{\epsfxsize=4.1in\epsfbox{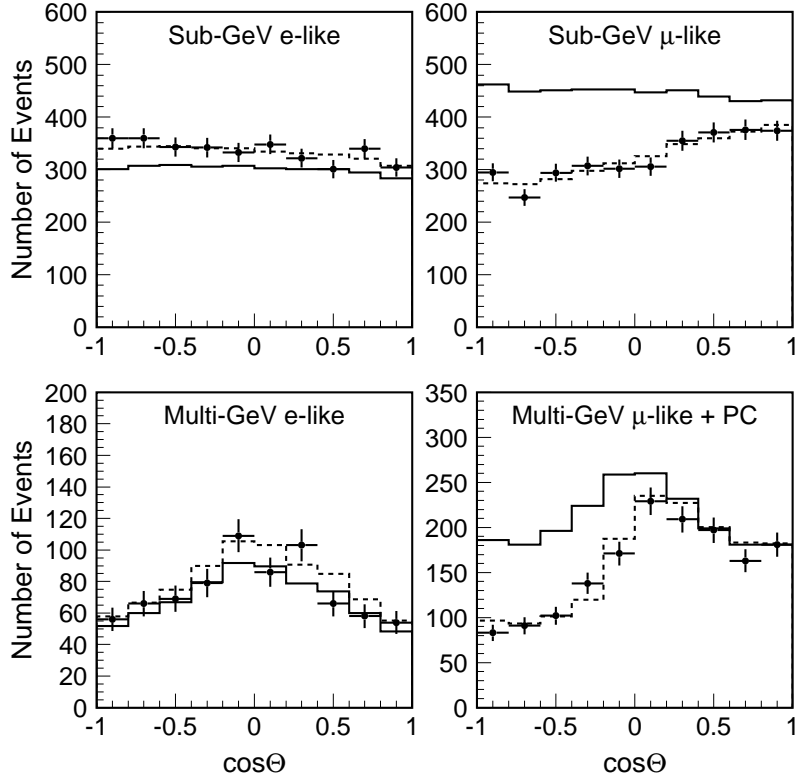}}   
\caption{Fluxes of atmospheric $\nu_e$ and $\nu_\mu$ as a function of
  zenith angle, as measured by Super-Kamiokande.  The solid lines show the no
  oscillation prediction, while the dashed line passing through the
  data points is the best-fit oscillation prediction.
\label{fig:sk_atmos}}
\end{figure}
As seen in Figure~\ref{fig:sk_atmos}, the deficit between the expected
and measured number of $\nu_\mu$ is largest at low energy and at
negative $\cos \theta$ (upward-going events).\cite{sk_atmos2}  This
dependence on energy and on the distance travelled by the neutrino is
characteristic of neutrino oscillations, and excludes a simple
normalization error.  These results were the first to establish
conclusively that atmospheric neutrinos oscillate.  The oscillation
seems to be of the type $\nu_\mu \to \nu_\tau$.  The atmospheric
neutrino effect has been confirmed by a number of other
experiments.\cite{other_atmos}

\begin{figure}[ht]
\centerline{\epsfxsize=4.1in\epsfbox{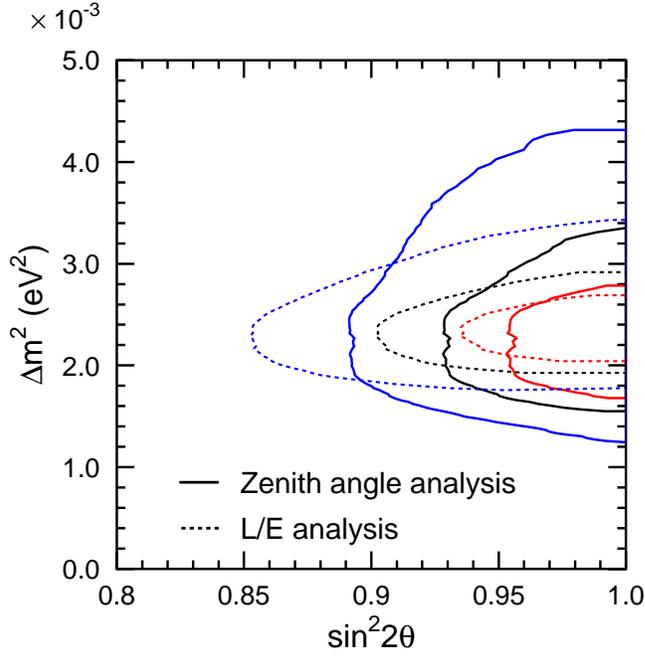}}   
\caption{Super-Kamiokande atmospheric neutrino mixing contours.  The
  two sets of contours are the 68\%, 90\%, and 99\% contours from two
  different analysis techniques.
\label{fig:sk_contours}}
\end{figure}

Figure~\ref{fig:sk_contours} shows the inferred mixing parameters from
fitting a two-flavour oscillation model to the atmospheric neutrino
data.\cite{sk_atmos}  The data favour $\Delta m^2 \approx 2.5\times
10^{-3}$~eV$^2$ and, surprisingly, a maximal mixing angle of $\theta
\approx 45^\circ$.  (The term ``maximal mixing'' refers to the fact
that each flavour eigenstate contains equal proportions of
the two mass eigenstates if $\theta=45^\circ$.)

\subsection{Long-Baseline Neutrino Oscillation Experiments}

Just as solar neutrino oscillations have been confirmed with
terrestrial (anti)-neutrinos by KamLAND, atmospheric neutrino
oscillations have recently been confirmed by the K2K long-baseline
neutrino oscillation experiment.\cite{k2k}  K2K produced a collimated
beam of $\nu_\mu$ by colliding a 12~GeV proton beam with an aluminum
target, thereby producing $\pi^+$'s.  These pions were then collected
and focused with a set of magnetic horns, and the collimated pion beam
then decayed in a long evacuated decay pipe by $\pi^+ \to \nu_\mu
\mu^+$.  The mean neutrino energy was 1.3~GeV, and the beam was aligned
with the direction of the Super-Kamiokande detector, located 250~km
away.  A set of near neutrino detectors measured the neutrino beam's
energy spectrum, interaction, and relative cross sections at a point
300~m from the pion production target.  By comparing the neutrino
energy spectrum and rate at the near detector to those measured at
Super-Kamiokande, the effects of neutrino oscillation over the 250~km
baseline can be inferred.  If the atmospheric neutrino effect is
really explained by neutrino oscillations, then K2K should see an
apparent ``disappearance'' of $\nu_\mu$, which oscillate into
$\nu_\tau$ that are too low in energy to be detected in Super-K
through charged current interactions.

Data collected by K2K between 1999 and 2004 in fact show a deficit of
muon-like events, and some indication of an energy dependence to the
$\nu_\mu$ disappearance effect as predicted for neutrino
oscillations.\cite{k2k}  A combined maximum likelihood fit to the
spectrum and rate excludes the null hypothesis of no oscillations at
the 4.0$\sigma$ level.  The best-fit oscillation parameters are
$\Delta m^2 = 2.8 \times 10^{-3}$~eV$^2$ and $\sin^2 2\theta=1$, which
are in excellent agreement with the values inferred from the
atmospheric neutrino data.

The MINOS experiment is a conceptually similar long-baseline
experiment in the United States.  MINOS uses the NUMI neutrino beam
produced by Fermilab's Main Injector, with a far detector located
$\sim 730$~km way in the Soudan mine in northern Minnesota, to study
oscillations of $\nu_\mu$.  MINOS should confirm K2K's results with
somewhat higher statistics, and at the time of writing results are
expected imminently\footnote{As this paper went to press the MINOS
collaboration released its first results, which confirmed $\nu_\mu$
disappearance in the NUMI beamline with $\Delta m^2 \approx 3 \times
10^{-3}$~eV$^2$ (publication pending).  }.

\subsection{The Three-Flavour Picture}

In the previous sections, the solar and atmospheric neutrino
oscillation effects were each analyzed separately in terms of
oscillations between two neutrino mass eigenstates.  In reality,
we know there are (at least) three flavour eigenstates, and so three
mass eigenstates.  Properly speaking we need to consider the
3$\times$3 MNS matrix, completely analogous to the CKM matrix for
quarks, which can be parameterized as:
\begin{equation}
\begin{array}{lccc}
U~= & \left( \begin{array}{ccc}
1 & 0 & 0 \\
0 & c_{23} & s_{23} \\
0 & -s_{23} & c_{23} \\
\end{array}
\right) &
\left( \begin{array}{ccc}
c_{13} & ~0~ & e^{i\delta} s_{13} \\
 0 & 1 & 0 \\
-e^{-i\delta} s_{13} & 0 & c_{13} \\
\end{array}
\right) &
\left( \begin{array}{ccc}
c_{12} & s_{12} & 0 \\
-s_{12} & c_{12} & 0 \\
0 & 0 & 1 \\
\end{array}
\right)
\end{array}
\label{eq:3x3nu}
\end{equation}
Here $c_{ij} \equiv \cos \theta_{ij}$ and $s_{ij} \equiv \sin
\theta_{ij}$.  

The $\theta_{12}$ term in this parameterization of the MNS matrix is
that which controls solar neutrino oscillations, which involve the
first and second mass eigenstates.  Experimentally $\theta_{12}
\approx 32^\circ$.\cite{sno_nsp}  For comparison, the equivalent
angle in the CKM matrix is the Cabibbo angle, which has the value
$\theta_C \approx 13^\circ$.  The mixing between the first and second
generations of leptons is thus much larger than the mixing between the
quark generations.  Similarly, $\theta_{23}$, which determines the
amplitude of atmospheric neutrino oscillations, is consistent with
maximal mixing ($\theta_{23} \approx 45^\circ$), even though its quark
counterpart equals just $\sim 2^\circ$!  It is unknown at present by
how much $\theta_{23}$ actually deviates from maximal mixing angle, or
whether this value is indicative of some kind of flavour symmetry
between the second and third generations.

By comparison, the middle part of Equation~\ref{eq:3x3nu} is poorly
constrained.  Limits on oscillations of reactor neutrinos at short
baselines ($\sim 1$~km) tell us that $\theta_{13} <
9^\circ$.\cite{chooz}  In fact, current measurements of $\theta_{13}$
are consistent with zero.  Presently nothing is known about the
complex phase $\delta$ in the MNS matrix, which if non-zero would
result in different oscillation patterns for neutrinos than for
antineutrinos.  This latter topic is of considerable interest.
Recalling that all observed instances of CP violation in physics can
be explained by a single complex phase in the CKM matrix, it is
exciting to realize that the observation that neutrinos oscillate
implies the possible existence of an entirely new source of CP
violation---one involving leptons rather than quarks!

\begin{figure}[th]
\centerline{\epsfxsize=3.6in\epsfbox{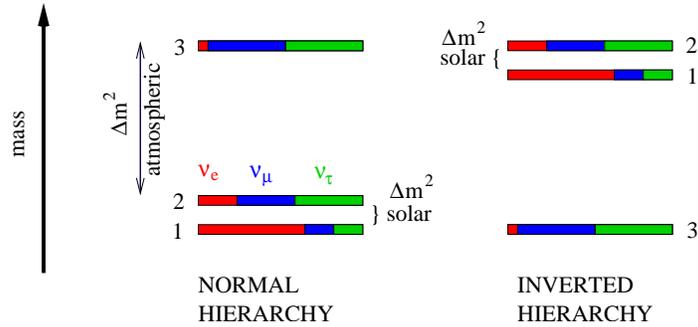}}   
\caption{Normal and inverted neutrino mass hierarchies.
\label{fig:hier}}
\end{figure}

Measurements of atmospheric and solar neutrino oscillations also
provide a partial determination of the pattern of the neutrino masses.
Solar and reactor neutrino data have determined that $\Delta m^2_{21}
\equiv m^2_2 - m^2_1 \approx 8.0 \times 10^{-5}$~eV$^2$ (see
Figure~\ref{fig:solarmixing})\cite{sno_nsp}, while atmospheric and
long baseline neutrino experiments\cite{sk_atmos,k2k} fix $|\Delta
m_{32}^2| \approx 2.5 \times 10^{-3}$~eV$^2$.  The solar neutrino
experiments have successfully inferred the sign of $\Delta m^2_{21}$
because the sign of the MSW effect in the Sun, which dominates in
solar neutrino oscillations, depends on the sign of $\Delta m^2$.  The
atmospheric neutrino data however has no significant sensitivity at
present to matter effects, and therefore it is not known whether $m_2
< m_3$ or rather $m_2 > m_3$.  The result is that there are two
possible mass hierarchies for the neutrino mass eigenstates.  The
so-called ``normal'' hierarchy has two light states and one heavier
state, with $m_1 < m_2 < m_3$, while in the ``inverted'' hierarchy
$m_3$ is the lightest state, with $m_1$ and $m_2$ being almost
degenerate in mass (see Figure~\ref{fig:hier}).  Note that neutrino
oscillation experiments are sensitive only to {\em differences} in
$m^2$, and do not measure the absolute mass scale, although lower
limits on the neutrino masses can be obtained by assuming the mass of
the lightest mass eigenstate to be zero.

\subsection{The LSND Result and the MiniBooNE Experiment}

Until this point I have put off discussion of one other neutrino
oscillation result that must be addressed.  The LSND collaboration
has reported 3.8$\sigma$ evidence for $\bar{\nu}_\mu \to \bar{\nu}_e$
oscillations of $\bar{\nu}_\mu$ produced from the decay of stopped
$\mu^+$ in a beam dump, over a propagation distance of $\sim
30$~meters.\cite{lsnd}  Other experiments, notably the KARMEN
experiment, have failed to confirm this effect\cite{karmen}, but do
not rule out the entire range of mixing parameters allowed by the LSND
result.  Mixing parameters with $\sin^2 2\theta \approx
10^{-3}-10^{-2}$ and $\Delta m^2 \sim 0.1-1$~eV$^2$ are consistent
with all data.\cite{lsnd_karmen}

The inferred value of $\Delta m^2$ from the LSND result is much larger
than those seen in solar and atmospheric neutrino experiments.  If
there are only three light neutrinos, then one can only form two
independent mass differences $\Delta m^2$.  If the LSND effect is due
to neutrino oscillation, then it implies a third independent value of
$\Delta m^2$, and so requires a fourth neutrino mass eigenstate.
However, the LEP measurements of the $Z$ boson's invisible decay width
confirm that there are only three active light neutrinos.\cite{lep}  A
fourth light neutrino, if it exists, must be sterile!  Even worse, more
detailed analyses of solar and atmospheric neutrinos show no
indication of any sterile neutrino admixtures, and are difficult to
reconcile with the existence of a single sterile
neutrino.\cite{maltoni}  By adding more than one sterile flavour,
enough wiggle room can be introduced to explain all of the oscillation
results.

The LSND result presents a particular problem for neutrino physics.
Because this result has not yet been confirmed by an independent
experiment, and because it has relatively drastic consequences such as
implying the existence of one or more sterile neutrino flavours, there
is widespread skepticism regarding its correctness.  That being said,
no fundamental flaw in the LSND experiment has been demonstrated, and
it is very possible that the result is correct.  Neutrinos may then be
more bizarre than anyone would have guessed!  At present the MiniBooNE
experiment at Fermilab is attempting to definitively check the LSND
result\cite{miniboone}, and is expected to produce first results for
$\nu_\mu \to \nu_e$ oscillations sometime in 2006.  Because the LSND
result has not yet been confirmed and cannot easily be accommodated
within the standard 3-flavour oscillation model, it is most often
ignored.  Only more data can determine whether it can be ignored
without great peril.

\section{Future Directions In Neutrino Oscillation}

In less than a decade we have evolved from a situation in which we had
no direct evidence that neutrinos oscillate to the present day, in
which both $\Delta m^2$ parameters are known to $\sim$10-20\%, and two
of the three neutrino mixing angles are known at least
approximately. One obvious way to proceed is to complete our picture
of the MNS matrix by attempting to measure the unknown mixing
parameters $\theta_{13}$ and $\delta_{CP}$, along with the sign of
$\Delta m_{32}^2$ that determines whether neutrinos have a normal or
inverted mass hierarchy.

\subsection{Measuring $\theta_{13}$, The Mass Hierarchy, and CP
  Violation At Long-Baseline Experiments}

The Super-K and K2K oscillation results seem to be of the type
$\nu_\mu \to \nu_\tau$, and are well described by a two-flavour mixing
model.\cite{sk_atmos,k2k}  However, in the full 3$\times$3 mixing
picture, there should be some probability that $\nu_\mu$'s will
instead oscillate into $\nu_e$'s in these experiments.  For an $L/E$
value tuned to $\Delta m^2_{32}$, this probability is given
by\cite{jhfloi}:
\begin{equation}
P(\nu_\mu \to \nu_e) \approx \sin^2 2\theta_{13} \sin^2 \theta_{23} \approx
\frac{1}{2} \sin^2 2\theta_{13}  
\label{eq:nueappear}
\end{equation}
Current limits on $\theta_{13}$ bound this probability to $<5\%$.

Because atmospheric neutrinos contain a significant fraction of
$\nu_e$, observing the small $\nu_\mu \to \nu_e$ transition
probability is not feasible.  Long baseline experiments however can
produce almost 100\% pure beams of $\nu_\mu$.  By searching for the
appearance of a small $\nu_e$ component in the beam at the oscillation
maximum, the value of $\theta_{13}$ may be inferred.

Equation~\ref{eq:nueappear} is only approximate, and the true $\nu_e$
appearance probability is modified by other mixing parameters and by
matter effects.  In particular, it can be shown that at the first
oscillation maximum, the $\nu_e$ appearance probability in vacuum is
altered in the presence of matter according to\cite{numiloi}:
\begin{equation}
P_{matter}(\nu_\mu \to \nu_e) \approx \left( 1+2\frac{E}{E_R} \right)
P_{vacuum}(\nu_\mu \to \nu_e)  
\label{eq:matter}
\end{equation}
where $E_R$ is a resonance energy given by $E_R = \Delta
m_{32}^2/(2\sqrt{2}G_FN_e)$.  This matter effect depends on the number
density of electrons $N_e$, and also on the magnitude and the sign of
$\Delta m_{32}^2$.  This matter effect correction is more significant
at large $L$ or $E$ values, and has the opposite sign for neutrinos
and antineutrinos.

A second confounding effect comes from the CP-violating phase of the
MNS matrix.  CP symmetry requires that neutrinos and antineutrinos
oscillate identically, so that $P(\nu_\mu \to \nu_e)=P(\bar{\nu}_\mu
\to \bar{\nu}_e)$ in vacuum.  However, a non-zero value of $\delta_{CP}$ can
make these probabilities unequal.  One can then define a CP asymmetry
for $\nu_e$ appearance which, ignoring matter effects, is given
by\cite{jhfloi}: 
\begin{equation}
\mathcal{A}_{CP} =  \displaystyle\frac{P(\nu_\mu \to \nu_e) - P(\bar{\nu}_\mu \to
\bar{\nu}_e)}{P(\nu_\mu \to \nu_e) + P(\bar{\nu}_\mu \to \bar{\nu}_e)}
 \simeq  \displaystyle\frac{\Delta m^2_{21} L}{4 E_\nu} \cdot \frac{\sin
2\theta_{12}}{\sin \theta_{13}} \cdot \sin \delta_{CP}
\label{eq:cp}
\end{equation}
The CP effect both changes $P(\nu_\mu \to \nu_e)$ and creates a
non-zero $\mathcal{A}_{CP}$.  Notice that the size of $\mathcal{A}_{CP}$ as measured at
the oscillation peak for the atmospheric neutrino $\Delta m_{32}^2$
depends on the solar neutrino parameters $\Delta m^2_{21}$ and
$\theta_{12}$ as well.  The reason for this is that, just as in the
quark sector, CP violation in the neutrino sector is an interference
effect: in this case, an interference between oscillations at the
solar and atmospheric frequencies.  To observe this effect,
oscillations at both $\Delta m^2$ values must be of roughly comparable
size, and $\theta_{13}$, which has the effect of coupling the
atmospheric and solar oscillations in Equation~\ref{eq:3x3nu}, must be
non-zero.  Fortunately for those of us interested in actually
observing CP violation by neutrinos, recent solar neutrino results
establishing the LMA solution imply that both solar mixing parameters
are reasonably large relative to the atmospheric neutrino mixing
parameters.  If $\theta_{13}$ is not too small, then observation of
non-zero $\mathcal{A}_{CP}$ may be possible.

\begin{figure}[t]
\centerline{\epsfxsize=3.5in\epsfbox{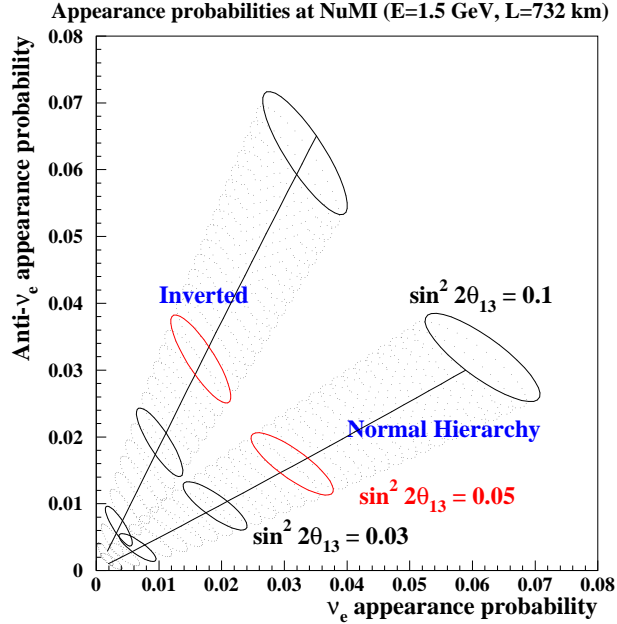}}   
\caption{Oscillation probabilities for $\nu_\mu \to \nu_e$
  vs. $\bar{\nu}_\mu \to \bar{\nu}_e$ for an off-axis experiment in
  the NUMI beamline.  The solid diagonal lines correspond to
  $\delta_{CP}=0$. 
\label{fig:numi_ellipses}}
\end{figure}

Because the $\nu_\mu \to \nu_e$ oscillation probability depends on
$\theta_{13}$, sign($\Delta m_{32}^2$), and $\delta_{CP}$, multiple
measurements at different energies and/or baselines will be needed to
disentangle the different effects.  Figure~\ref{fig:numi_ellipses}
illustrates the dependence of $P(\nu_\mu \to \nu_e)$ and
$P(\bar{\nu}_\mu \to \bar{\nu}_e)$ on the different oscillation
parameters, for monoenergetic (anti)neutrino beams with $E=1.5$~GeV
and $L=732$~km.  The sign of $\Delta m_{32}^2$ defines the normal and
inverted mass hierarchies, dividing the predicted probabilities into two
separate ``cones''.  Increasing $\theta_{13}$ moves one out along
either cone to larger oscillation probabilities.  With $\theta_{13}$
and sign($\Delta m_{32}^2$) fixed, varying $\delta_{CP}$ traces out an
ellipse in the plane, as shown in the figure.  A suitably precise
measurement of the neutrino and antineutrino appearance probabilities
could determine the mass hierarchy for largish $\theta_{13}$, as well
as defining an allowed region in the $\theta_{13}-\delta_{CP}$ plane.
Measurements at different choices of $L$ and $E$ will have different
sensitivity to matter effects and to $\delta_{CP}$ (see
Equations~\ref{eq:matter} and \ref{eq:cp}), and can be used to break
any remaining parameter degeneracies.

At present just one experiment to study $\nu_e$ appearance at the
atmospheric $\Delta m^2$ has been funded.  This is the T2K experiment
in Japan.\cite{jhfloi}  T2K will use a megawatt-scale proton beam at
the Japan Proton Accelerator Research Complex (JPARC) in Tokai to
produce a $\nu_\mu$ beam that will be directed towards
Super-Kamiokande, 295~km away.  By pointing the neutrino beam about
2$^\circ$ away from Super-K, T2K will take advantage of a trick called
``off-axis focusing'', which results in a nearly monoenergetic
neutrino beam with a peak energy of $\sim$700~MeV.  At these energies
the dominant interactions are charged current quasi-elastic ($\nu_\ell
+ n \to p + \ell$).  A set of sophisticated near detectors will
measure the beam properties before oscillation.  T2K will have
approximately 50 times greater statistics than K2K.  With its
relatively low beam energy and small baseline, T2K is relatively
insensitive to matter effects.

The most important backgrounds to $\nu_e$ appearance at T2K are a
small component of $\nu_e$ in the beam itself, and neutral current
$\pi^0$ production at Super-K.  The latter is only a background to
$\nu_e$ appearance if Super-K fails to detect one of the two
$\gamma$-rays.  This could happen in very asymmetric decays in which
one photon takes the bulk of the $\pi^0$'s energy, or if optical
scattering of Cherenkov light sufficiently obscures one of the two
Cherenkov rings.  For five years of running at nominal luminosity ($5
\times 10^{21}$ protons on target), T2K expects to achieve sensitivity
to $\theta_{13}$ down to $\sin^2 2\theta_{13} \approx 10^{-2}$ (the
exact limit depends on the value of $\delta_{CP})$.\cite{t2kcdr}  The
measured value of $\theta_{13}$ is partially degenerate with
$\delta_{CP}$, and separating the two parameters will require
additional measurements with antineutrinos and/or at other baselines.
Assuming that T2K successfully detects $\nu_e$ appearance in the
$\nu_\mu$ beam, the natural followup is to switch the polarity of the
beam and look for $\bar{\nu}_\mu \to \bar{\nu}_e$.  With a beam power
upgrade and possibly the construction of a larger far detector, this
``phase 2'' program could then begin to explore CP violation in the
neutrino sector.

In addition to measuring the $\nu_e$ appearance probability,
future long-baseline neutrino experiments such as T2K will measure the
$\nu_\mu$ disappearance probability with much higher statistics,
allowing precision measurements of $\Delta m_{32}^2$ and
$\theta_{23}$.  Such measurements can test how close $\theta_{23}$ is
to maximal mixing ($45^\circ$), explore whether any fraction of the
$\nu_\mu$ flux is oscillating to a non-interacting (sterile) neutrino
flavour, and test the energy dependence of the neutrino oscillation
prediction with high precision.

Although T2K is currently the only funded new long-baseline experiment to
search for $\nu_e$ appearance, the NO$\nu$A collaboration in the US
has proposed building a new off-axis detector, optimized for detecting
electron appearance, in Fermilab's NUMI beamline.\cite{numiloi}  At a
baseline of $\sim 730$~km and a beam energy of $\sim 2$~GeV, the
NO$\nu$A experiment could have some sensitivity to matter effects and
the sign of the mass hierarchy if $\theta_{13}$ is not too small, and
would otherwise have similar sensitivity to $\nu_e$ appearance as T2K.
The proposed far detector is a massive finely segmented liquid
scintillator detector.  The NO$\nu$A proposal is currently in the
early stages of the approval process.

\subsection{Reactor Neutrino Experiments}

An alternate approach to measuring $\theta_{13}$ is to do precision reactor
neutrino experiments at short baselines.  The full 3-flavour formula
for reactor $\bar{\nu}_e$ oscillation is\cite{reactor13}:
\begin{equation}
\begin{array}{ll}
P(\bar{\nu}_e \to \bar{\nu}_e) = 1 & -
\sin^22\theta_{13} \sin^2 \left( \frac{1.27 \Delta m_{13}^2 L}{E}
\right)\\
& - \cos^4 \theta_{13} \sin^22\theta_{12} \sin^2 \left( \frac{1.27 \Delta m_{21}^2 L}{E} \right)
\end{array}
\end{equation}
The first term, which is proportional to $\sin^22\theta_{13}$ and
depends on the larger $\Delta m^2$ value, dominates over the second at
short baselines.  The second term only becomes significant at reactor neutrino
energies for $L \approx 200$~km.  KamLAND was successfully able to use
the second term to confirm the solar neutrino effect\cite{kamland},
but experiments at shorter baselines instead yield limits on
$\theta_{13}$.  Currently the best limits on $\theta_{13}$ come from
the CHOOZ reactor neutrino experiment, which limits $\sin^2
2\theta_{13} < 0.15$ at the 90\% C.L.\cite{chooz}

A new reactor neutrino experiment with high statistics and improved
systematics may be able to achieve significantly improved
$\theta_{13}$ sensitivity.\cite{reactor13}  The keys to better
sensitivity are to use a very intense reactor, with power in the gigawatt
range, and to use both a near detector right next to the reactor and a
far detector 1 or 2~km away in order to cancel systematics between the
two detectors.  A significant advantage of reactor $\theta_{13}$
experiments is that they are not sensitive to CP-violating effects
(which can only be measured in an appearance measurement, not in a
disappearance measurement), nor to matter effects, which are
negligible at the relevant $L$ and $E$ values.  A good reactor
neutrino experiment therefore would provide a clean measurement of
just $\theta_{13}$.  This provides significant complementarity to
long-baseline $\nu_e$ appearance experiments, which are sensitive to a
combination of $\theta_{13}$, the mass hierarchy, and $\delta_{CP}$.

An added advantage of reactor $\theta_{13}$ experiments is that they
are relatively inexpensive, with a typical estimated price tag of
$\sim$\$50M.  For this reason, it seems that experimenters have
proposed new experiments at virtually every reactor complex in the
world with significant power output.  Prominent sites for proposed
experiments include Daya Bay in China, Braidwood in Illinois, and the
Double CHOOZ proposal in France, although this list is far from
exhaustive.\cite{reactor13} It seems likely that one or more of these
proposals will be funded, but at present it is not clear which ones.
The physics case for a sensitive reactor $\theta_{13}$ experiment seems
compelling, however.

\section{Altering The Standard Model To Accommodate Neutrino Mass}

In the Standard Model, neutrinos have zero mass.  This is not simply
an {\em ad hoc} assumption, but a consequence of the fact that the
Standard Model does not contain right-handed neutrino fields.  Fermion
mass terms in the Standard Model Lagrangian have the form $-m\bar{\psi}\psi =
-m(\bar{\psi_L}\psi_R + \bar{\psi_R}\psi_L)$.  Without a right-handed
field, no such term can exist.  In this section I shall examine
possible ways in which the Standard Model may be extended to include
non-zero neutrino mass.

\begin{figure}[t]
\centerline{\epsfxsize=3.1in\epsfbox{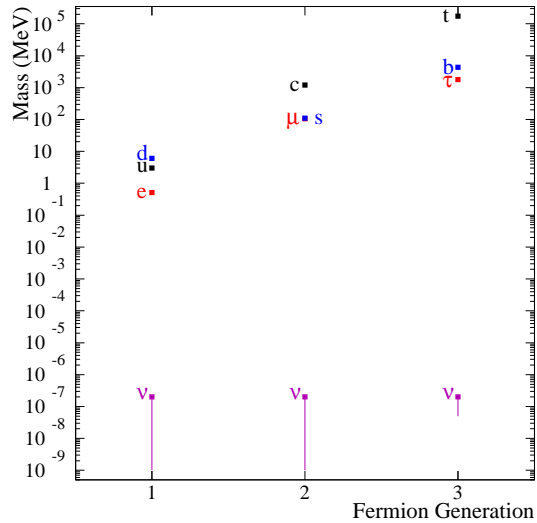}}   
\caption{Masses of the Standard Model fermions.  The purple lines
  indicate the range of allowed neutrino masses for $\nu_1$, $\nu_2$,
  and $\nu_3$, assuming a normal mass hierarchy.
\label{fig:fermion_masses}}
\end{figure}

The most obvious solution to this problem is to simply add
right-handed neutrino states $\nu_R$ to the Standard Model and to give
them Yukawa couplings to $\nu_L$ through the Higgs field, just like
other fermions.  This is called a Dirac mass term.  While
superficially this places neutrinos on the same footing as the other
fermions, one striking difference is that $\nu_R$, having neither
charge, colour, nor couplings to $W^\pm$ or $Z$, are sterile fields
(i.e. they don't couple to the vector gauge bosons)---making them in
an important sense unlike all other Standard Model particles.  An
additional puzzle is that in order to explain the smallness of
neutrino masses, their Yukawa couplings must be made anomalously
small.  As illustrated in Figure~\ref{fig:fermion_masses}, within each
generation the charged fermions are separated in mass by no more than
1 or 2 orders of magnitude, but the neutrino mass eigenstates are many
orders of magnitude lighter than their charged counterparts.  While it
may rightfully be objected that we have no good explanations for the
numerical values of the masses of {\em any} fermions, the disparity
between neutrino and charged fermion masses suggests that neutrinos
might not simply acquire mass in the same manner as other fermions.

Another possible way to add neutrino mass terms is to recognize that
there already exists a right-handed neutral fermion in the Standard
Model---namely, the antineutrino.  Is it possible to identify $\nu_R$
with the antineutrino, and so generate mass terms of the form
$\bar{\nu_L}\nu_R$ by combining a neutrino with its antineutrino?  For
charged fermions, the answer would clearly be no: since particles and
antiparticles have opposite charges, a term that directly couples a
fermion to its antifermion violates charge conservation!  However, the
situation is different with neutrinos, which are chargeless particles.
The Majorana neutrino hypothesis takes advantage of this
chargelessness by positing that an antineutrino is just a neutrino
with its spin flipped by $180^\circ$!  One might then form a ``Majorana
mass'' term that couples a left-handed neutrino with the right-handed
antiparticle.


Nonetheless, within the minimal Standard Model, Majorana mass terms
are in fact forbidden. The reason is that although the Standard Model
does not conserve either baryon number $B$ or lepton number $L$
non-perturbatively, it does conserve the quantity $B-L$ exactly.  A
Majorana mass term on the other hand results in $|\Delta(B-L)|=2$.  It
turns out that without extending the Standard Model particle content
in some manner, a $B-L$ violating term cannot be generated at any
order, even as an effective operator.\cite{akhmedov}

However, the addition of an {\em additional} right-handed Majorana
field to the Standard Model can resolve the problem.  Let $\nu_L$ be a
2-component field describing left-handed neutrino/right-handed
antineutrino that couples to weak interactions.  Let $\nu_R$ now
denote an {\em additional} right-handed Majorana field, independent of
$\nu_L$, which does not couple to weak interactions.  Because $\nu_R$
is an electroweak singlet, it can possess a bare Majorana mass term
that couples $\nu_R$ to its antiparticle.  We may also have a Dirac
mass term (Yukawa coupling) between $\nu_R$ and the active light
neutrino $\nu_L$.  The mass terms in the Lagrangian are
then\cite{elliott,akhmedov}:
\begin{equation}
-\Delta \mathcal{L} = m_D \bar{\nu_L}\nu_R + \frac{1}{2}m^*_R\nu_R^TC\nu_R + h.c.  
\label{eq:mass_lagrangian}
\end{equation}
The first term here is a Yukawa coupling between $\nu_L$ and $\nu_R$,
and is referred to as a Dirac mass term.  For charged fermions, this
is the only allowed mass term.  The second term is a Majorana mass
term that couples $\nu_R$ to its antiparticle.  This term is allowed,
and violates no gauge symmetries, provided that $\nu_R$ is
chargeless---that is, that $\nu_R$ is its own antiparticle.  It's
evident that $\nu_L$ and $\nu_R$ should be thought of here as separate
fields, with independent mass terms and in fact different masses.

Having written down Equation~\ref{eq:mass_lagrangian}, some magic now
results.  We can rewrite the mass term in the Lagrangian as
\begin{equation}
-\Delta \mathcal{L}  = \frac{1}{2} (\nu_L, \nu_R)
\left(
\begin{array}{cc}
0 & m_D \\
m_D & m_R \\
\end{array}
\right)
\left(
\begin{array}{c}
\nu_L \\
\nu_R \\
\end{array}
\right)
\label{eq:seesaw_matrix}
\end{equation}
Equation~\ref{eq:seesaw_matrix}, which is not obviously diagonal, can
be diagonalized to yield the physical mass eigenstates.  There are two
eigenvalues:
\begin{equation}
M_{heavy} \approx m_R, \phantom{XXXXXXXX} 
M_{light} \approx \frac{m^2_D}{m_R}  
\end{equation}
Because $\nu_R$ is an electroweak singlet, its mass is not protected
by any electroweak symmetry, and the theoretical expectation is that
it should be quite massive---possibly at the GUT
scale.\cite{akhmedov,elliott}  On the other hand, we would naively
expect $m_D$ to be similar in size to the Dirac masses of other
fermions.  If we take $m_R = 10^{15}$~GeV as a typical GUT-scale
energy and $m_D =200$~GeV as representative of the Yukawa coupling of the
heaviest charged fermion, we would estimate the largest light neutrino
mass to be $M_{light} = (200~{\rm GeV})^2/(10^{15}~{\rm GeV}) =
0.04$~eV.  This value is exactly the right order of magnitude for the
neutrino mass inferred by $\sqrt{\Delta m^2_{32}} \approx 0.05$~eV!

Something semi-miraculous has occurred.  By introducing a right-handed
neutrino with a mass near the GUT scale, as is motivated by GUT
models, with a ``normal'' Dirac coupling $m_D$ to $\nu_L$, we
naturally produce very light neutrino masses for $\nu_L$, without
having to fine-tune the Dirac mass coupling.  The heavier that
$M_{heavy}$ is, the lighter that $M_{light}$ becomes, which gives rise
to the name ``seesaw mechanism'' for this method of generating light
neutrino masses.  Obviously the close numerical correspondence between
$\sqrt{\Delta m_{32}^2}$ and $M_{light}$ in the previous paragraph
should not be taken too seriously, since we do not know the exact
values of $m_R$ or $m_D$ to use in the calculation.  The exact mass
calculation in fact depends on the details of the physics at higher
energy scales.  Nonetheless, the seesaw mechanism provides as least a
proof of principle as to how very light neutrino masses can be
generated without fine-tuning the Dirac mass coupling, while providing
a fascinating example of a novel method of generating masses for
fundamental particles.

\section{Determining The Absolute Mass Scale Of Neutrinos}

Although neutrino oscillation measurements demonstrate the existence
of neutrino masses, they cannot determine the absolute values of the
masses, since oscillations are only sensitive to {\em differences} in
$\Delta m^2$.  One may make an educated guess of the absolute masses
if one assumes that each mass eigenstate is much heavier than the
previous one ($m_1 \ll m_2 \ll m_3$), reproducing the pattern of
charged fermion masses.  In this limit then $m_1 \approx 0$~eV, $m_2
\approx \sqrt{\Delta m_{21}^2} = 0.009$~eV, and $m_3 \approx
\sqrt{\Delta m_{32}^2} = 0.05$~eV.  For an inverted hierarchy we get
$m_3 \approx 0$~eV and $m_1 \approx m_2 \approx 0.05$~eV.

Of course it is not clear whether a strict mass hierarchy should hold.
As the mass of the lightest mass eigenstate increases, the three mass
eigenstates approach a limit of degenerate masses. The best upper
limits on neutrino mass come from cosmology.  Massive neutrinos act as
a form of hot dark matter that tends to wash out clustering at small
angular scales during structure formation, since relativistic dark
tends to ``stream out'' of small density perturbations, but not larger
ones.  This effect can leave signatures in the cosmic microwave
background radiation, in weak lensing surveys, and in large scale
structure surveys.  While the exact limits obtained depend on which
data sets are included in the fits and with what priors, the published
limits\cite{elgaroy} for the sum of the three mass eigenstates range from $\sim
0.4-0.7$~eV.  It would not be far wrong to say that
cosmology limits the mass of any individual mass eigenstate to be
$<0.2$~eV.

Less stringent but more model-independent limits come from
measurements of the energies of the products of weak decays.  Notable
among these are studies of the endpoint of tritium beta decay.  If
neutrinos have non-zero mass, this mass has the effect of reducing the
maximum energy available for the $\beta$ particle in the decay.
Careful measurements of the shape of the $\beta$ energy spectrum at
the endpoint limit the effective mass of a $\nu_e$ (the weighted
average of its mass eigenstates) to $<2.5$~eV at the 95\% C.L.  The
KATRIN collaboration has proposed a next generation tritium endpoint
measurement with sensitivity down to 0.2~eV, which might be able to
measure or rule out the case of three quasi-degenerate
masses.\cite{katrin}

\subsection{Neutrinoless Double Beta Decay}

In a class by themselves are experiments to measure neutrinoless
double beta decay.  Normal double beta decay is a doubly weak process
in which a nucleus decays by simultaneously emitting two electrons and
two $\bar{\nu}_e$.  Double beta decay can occur when single beta decay
is energetically forbidden, but the $|\Delta Z|=2$ process is
energetically allowed.  If neutrinos are Majorana particles (so that a
neutrino is its own antiparticle), then instead of emitting two
neutrinos, a Feynman diagram exists in which a virtual neutrino is emitted
then reabsorbed as an antineutrino.  The result is a beta decay in
which two electrons but no neutrinos are emitted.  Neutrinoless double
beta decay violates lepton number by $|\Delta L|=2$, and differs
kinematically from ordinary double beta decay in that the two emitted
electrons now contain all of the emitted energy of the transition.
The experimental signature of neutrinoless double beta decay is
therefore a peak right at the endpoint of the distribution of the sum
of the two electrons' energies.  

The rate of neutrinoless double beta decay depends on the available
phase space and on nuclear matrix elements of the decaying nucleus,
but can also be shown to depend an effective neutrino mass
by\cite{elliott}:
\begin{equation}
R \propto \langle m_\nu \rangle^2 = \left| \sum_{i}^N U_{ei}^2 m_i \right|^2  
\label{eq:0nu}
\end{equation}
The effective mass depends on the elements in the first row of the MNS
matrix.  The mass values enter because they control how much of the
``wrong'' chirality is mixed into each neutrino, determining the
transition of a Majorana neutrino into an antineutrino.  (Recall that
by the Majorana neutrino hypothesis an antineutrino is just a neutrino
of the opposite chirality.)

Positive detection of neutrinoless double beta decay would arguably be
the most exciting possible result in neutrino physics, since it would
simultaneously establish that neutrinos are Majorana particles, show
that lepton number is violated, and settle what the absolute values of
the neutrino masses are.  This phenomenon has been searched for in
many candidate nuclei, but no confirmed detections have been found.
The best upper limits comes from the $^{76}$Ge system, which limits
$\langle m_\nu \rangle < 0.35$~eV at the 90\% C.L.\cite{ge76}

While neutrinoless double beta decay experiments are tremendously
difficult due to the rarity of such decays and the existence of
various potential backgrounds, many proposals for next generation
experiments exist.  These proposals rely on much larger exposures
(kilograms of material $\times$ years of data-taking), and on
sophisticated active or passive means to reject backgrounds.  One such
proposed experiment is the MAJORANA experiment, whose goal is to
collect 2500~kg-years exposure of $^{76}$Ge to achieve sensitivity
down to $\langle m_\nu \rangle < 0.05$~eV.\cite{majorana}  The EXO
experiment will look for neutrinoless double beta decay in 10 tonnes
of $^{136}$Xe in a liquid or gas TPC, and will attempt to tag the
resulting barium ion using spectroscopic techniques to eliminate
backgrounds, with a sensitivity goal of $\sim 0.01$~eV.\cite{exo}
These and other next-generation experiments, if successful, have some
hope of covering the expected range for $\langle m_\nu \rangle$ for
degenerate neutrino masses and for the inverted hierarchy.  A null
result would tell us that neutrinos, if Majorana particles, must have
a normal mass hierarchy, but by itself could not distinguish between
the possibilities that neutrinos either have a normal mass hierarchy
or are simply not Majorana particles.  (In principle, though, a
determination from long baseline neutrino oscillation experiments that
neutrinos have an inverted mass hierarchy, combined with a null result
from a sufficiently sensitive neutrinoless double beta decay
experiment, could demonstrate that neutrinos are not Majorana
particles!)

\section{Conclusions}

The past decade of neutrino physics has been revolutionary.  We have
gone from having no confirmed evidence for neutrino physics beyond the
Standard Model to the current situation, in which oscillation has been
observed in four separate systems, with reasonably precise
measurements of two $\Delta m^2$ values and two of the four
independent mixing parameters in the MNS matrix (assuming that this
matrix really is unitary!)  Neutrino oscillation is new physics beyond the
Standard Model, and requires the addition of new fields and new
parameters to the Standard Model.  It may even point to the existence
of new mechanisms of mass generation.

With the discovery of neutrino mixing, we are now entering an era of
precision lepton flavour physics.  Just as the study of the CKM matrix
has been one of the most important areas in particle physics for
decades, studies of lepton flavour may lead to new insights into
the origins of flavour, CP violation, and the relationship between
quarks and leptons.  

In the near future, the experimental emphasis is likely to be on
determining $\theta_{13}$ through long baseline or reactor neutrino
experiments, as well as precisely testing the predictions of the
neutrino oscillation model.  Longer term we can aspire to looking for
CP violation by neutrinos in long baseline oscillation experiments,
searching for neutrinoless double beta decay in an attempt to answer
the Majorana vs. Dirac neutrino question, and improving limits on
neutrino mass from direct kinematic experiments or from cosmology.
All the while anomalies like the controversial LSND result remind us
that neutrinos may present other surprises that we have not even
anticipated yet.

Clearly I'm an optimist about the future of neutrino research.  Given
that neutrino oscillations are the first new particle physics beyond
the Standard Model, the (over?)abundance of new proposals for
experiments, and the fact that even today neutrino experiments are
probing new physics at a tiny fraction of the cost of large collider
experiments, how can I not be an optimist for the future of our field?
I hope in the end that the reader agrees with me in this regard.

\section*{Acknowledgments}
I wish to thank the organizers of the Lake Louise Winter Institute for
inviting me to speak at the Institute.  John Ng and Maxim Pospelov
provided valuable discussions about the theory of Majorana neutrino
masses, but should be held blameless for all of my mistakes in
presenting it.

\end{document}